\documentclass[aps,prb,preprint,groupedaddress,showpacs,amsmath,amssymb]{revtex4}

\usepackage{graphicx}
\usepackage{dcolumn}
\usepackage{bm}

\begin{document}

\title{Structural phase transition in Ba(Fe$_{0.973}$Cr$_{0.027}$)$_2$As$_2$  single crystals.}

\author{S. L. Bud'ko}
\author{S. Nandi}
\author{N. Ni}
\author{A. Thaler}
\author{A. Kreyssig}
\author{A. Kracher}
\author{J.-Q. Yan}
\author{A. I. Goldman}
\author{P. C. Canfield}
\affiliation{Ames Laboratory US DOE and Department of Physics and Astronomy, Iowa State University, Ames, IA 50011, USA}

\date{\today}

\begin{abstract}

We present thermodynamic, structural and transport measurements on Ba(Fe$_{0.973}$Cr$_{0.027}$)$_2$As$_2$ single crystals. All measurements reveal sharp anomalies at $\sim 112$ K. Single crystal x-ray diffraction identifies the structural transition as a first order, from the high-temperature tetragonal $I4/mmm$ to the low-temperature orthorhombic $Fmmm$ structure, in contrast to an earlier report.

\end{abstract}


\maketitle

The recent discoveries of superconductivity in Fe-As based materials, F-doped LaFeAsO \cite{kam08a} and K-doped BaFe$_2$As$_2$, \cite{rot08a} resulted in a large number of experimental and theoretical studies of the materials with similar structural motifs. The $AE$Fe$_2$As$_2$  ($AE$ = Ba, Sr, Ca) family of compounds soon became a model system for many studies of iron-arsenides, in part, due to the availability of large, high-quality single crystals of pure and doped materials and notable reproducibility of the results between different experimental groups. \cite{nin08c,chu09a,nin09a,fan09a} The parent compounds,  $AE$Fe$_2$As$_2$  ($AE$ = Ba, Sr, Ca), were shown to exhibit a coupled, structural/antiferromagnetic phase transition, all with the transition temperatures above 100 K. Structurally, in all three parent compounds, the high temperature, tetragonal (space group $I4/mmm$) symmetry changes to the lower temperature, orthorhombic one (space group $Fmmm$) at this transition. \cite{rot08b,yan08a,nin08a} It has been shown that (although the transition temperature decreases, and, in some cases, the structural and magnetic transitions split) for several types/sites of doping, e.g. Sn incorporated in BaFe$_2$As$_2$ crystals as a result of the use of Sn flux, \cite{nin08b,suy09a} (Ba$_{1-x}$K$_x$)Fe$_2$As$_2$, \cite{rot08c} and Ba(Fe$_{1-x}$Co$_{x}$)$_2$As$_2$, \cite{pra09a,les09a} the nature of the structural phase transition ($I4/mmm$ to $Fmmm$ on cooling) is very robust. With this in mind, the claim \cite{sef09a} that for small Cr doping, such as Ba(Fe$_{0.98}$Cr$_{0.02}$)$_2$As$_2$, the tetragonal to orthorhombic symmetry breaking is replaced by an $I4/mmm$ to $I4/mmm$ (tetragonal to tetragonal) transition with a decrease of both lattice parameters resulting in a volume reduction, was unexpected, exciting and, in our opinion, worth further, detailed studies. In addition to simply being anomalous, this difference could be important, since no superconductivity was reported in any of the Cr-doped BaFe$_2$As$_2$ samples. \cite{sef09a}
\\

Single crystals of Ba(Fe$_{0.973}$Cr$_{0.027}$)$_2$As$_2$ were grown out of self flux using conventional high-temperature solution growth techniques. \cite{nin08c,can92a}  Small Ba chunks, FeAs and CrAs powder were mixed together according to the ratio Ba:FeAs:CrAs = 1:3.9:0.1. The mixture was placed into an alumina crucible with a second, "catch", crucible containing quartz wool placed on top. Both crucibles were sealed in a quartz tube under a $\sim 1/3$ atmosphere of Ar gas. The sealed quartz tube was heated up to 1180$^\circ $C over 12 hours, held at 1180$^\circ$ C for 10 hours, and then cooled to 1050$^\circ$ C over 46 hours. Once the furnace reached 1050$^\circ$ C, the excess FeAs/CrAs liquid was decanted from the plate-like single crystals. Elemental analysis of the samples was performed by wavelength dispersive analysis (WDS) in a JEOL JXA-8200 electron microprobe. WDS measurements were made at a total of twenty locations on four Ba(Fe$_{1-x}$Cr$_{x}$)$_2$As$_2$ crystals from the batch used for all measurements in this work. The average $x$ value measured at these locations is 0.027, and the error bar, which is defined as two times the standard deviation of the $x$ values measured on these locations, is 0.002.  This is within the error bars of the $x = 0.02 \pm 0.01$ sample studied in Ref. \onlinecite{sef09a}. However, based on a comparison of the data presented below with the data in Ref. \onlinecite{sef09a}, it is likely that our sample has slightly more Cr (a slightly larger $x$-value) than $x = 0.02 \pm 0.01$, but significantly less than $x = 0.04 \pm 0.01$.

Anisotropic, temperature-dependent magnetic susceptibility and field-dependent magnetization were measured in a commercial, Quantum Design (QD) MPMS magnetometer. Measurements of ac (magneto)resistivity and Hall effect ($f = 16$ Hz, $I = 3 - 5$ mA) were performed using the ACT option of a QD PPMS instrument. Electrical contacts to the sample were made with Epotek H20E silver epoxy. A standard four-probe technique was used for resistivity. Hall resistivity data were collected in a four wire geometry, switching the polarity of the magnetic field ($H \| c$) to remove magnetoresistance components due to the slight misalignment of the voltage wires. Temperature-dependent Hall resistivity was measured in $H = 90$ kOe applied field. The heat capacity data on the samples were measured using a hybrid adiabatic relaxation technique of the heat capacity option in a QD PPMS instrument. Thermal expansion data were obtained using a capacitive dilatometer constructed of OFHC copper, mounted in a QD PPMS instrument. A detailed description of the dilatometer is presented elsewhere. \cite{sch06a}

Temperature dependent, single crystal X-ray diffraction measurements were performed on a four-circle diffractometer using Cu $K_{\alpha}$ radiation from a rotating anode X-ray source, selected by a germanium $(1~1~1)$ monochromator for high angular resolution. For the measurements, a plate like single crystal with dimensions of $4.0 \times 2.5 \times 0.7$ mm$^3$ was selected and attached to copper sample holder on the cold finger of a closed cycle, Displex refrigerator. The diffraction patterns were recorded while the temperature was varied between 25 K and 125 K. The mosaicity of the investigated Ba(Fe$_{0.973}$Cr$_{0.027}$)$_2$As$_2$ single crystal was 0.04 degrees full-width-at-half-maximum (FWHM) as measured from the rocking curve of the $(0~0~10)$ reflection.
\\

\begin{figure}
\begin{center}
\includegraphics[angle=0,width=120mm]{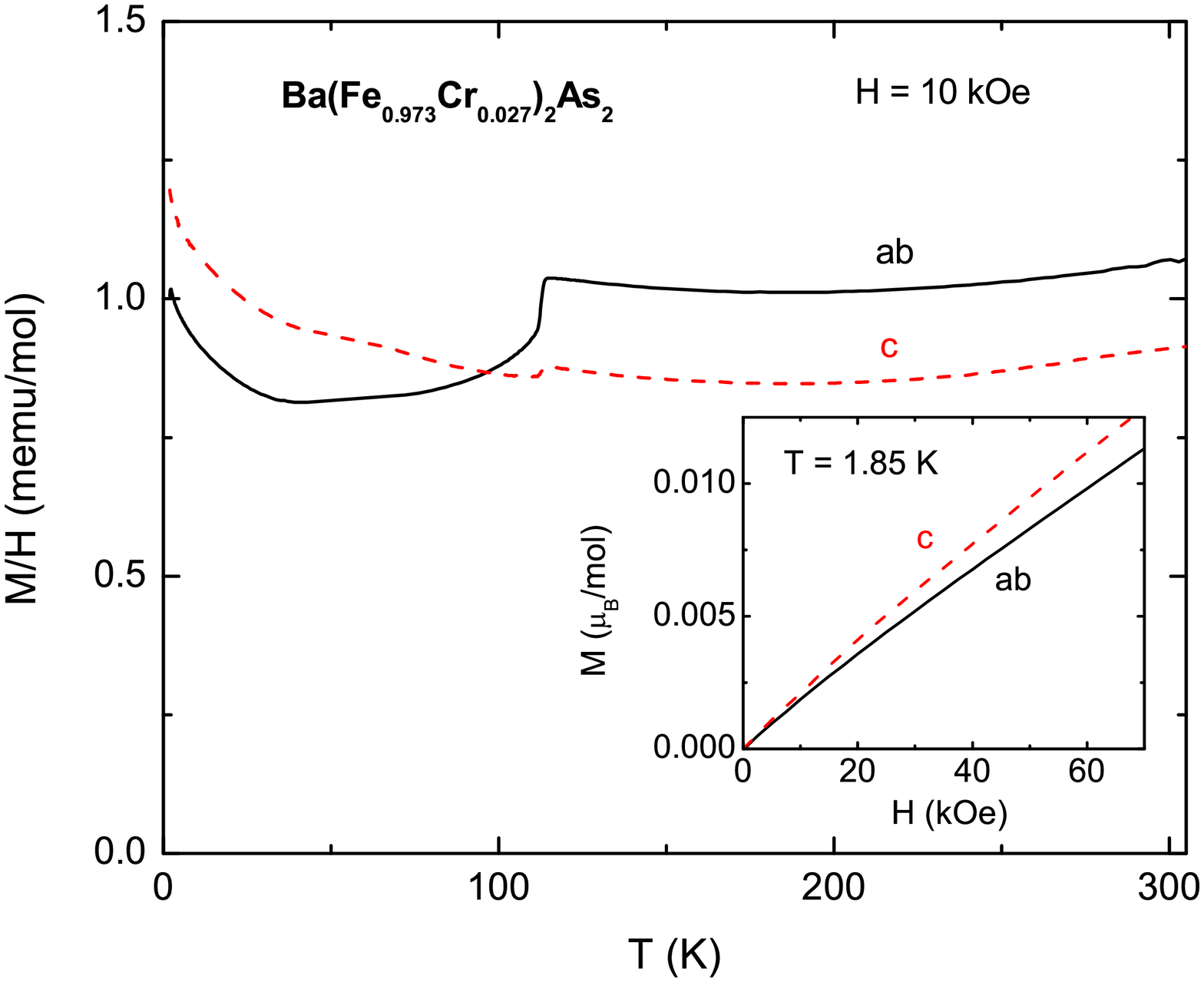}
\end{center}
\caption{(Color online) Anisotropic, temperature dependent susceptibility for Ba(Fe$_{0.973}$Cr$_{0.027}$)$_2$As$_2$ single crystals. Inset  shows anisotropic field dependent magnetization at $T =1.85$ K.}\label{F1a}
\end{figure}

\begin{figure}
\begin{center}
\includegraphics[angle=0,width=120mm]{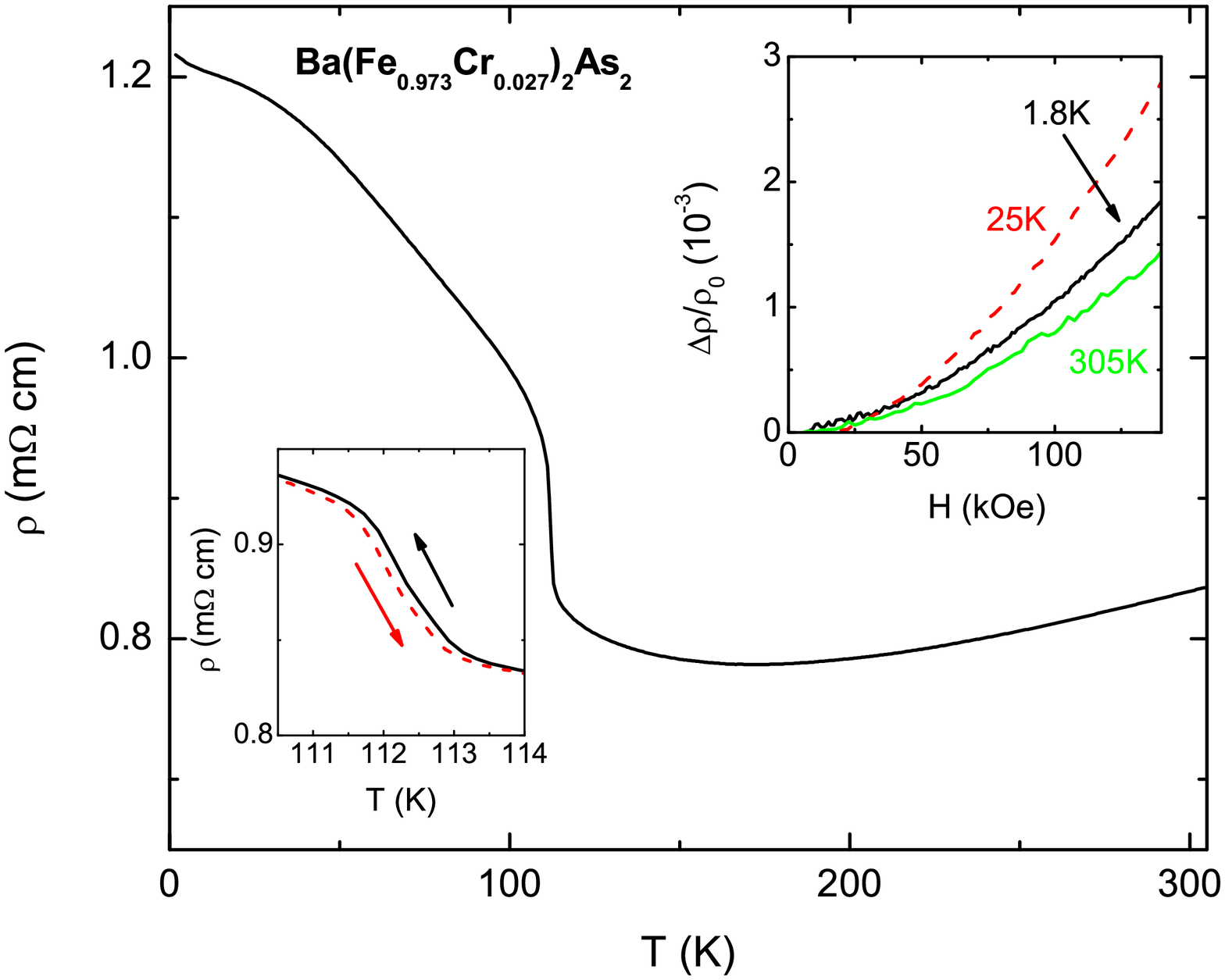}
\end{center}
\caption{(Color online) Temperature dependent resistivity for Ba(Fe$_{0.973}$Cr$_{0.027}$)$_2$As$_2$ single crystals. Insets show hysteresis at the phase transition (left) and magnetoresistivity for $H \| c$, $I \| ab$ (right). }\label{F1b}
\end{figure}

\begin{figure}
\begin{center}
\includegraphics[angle=0,width=120mm]{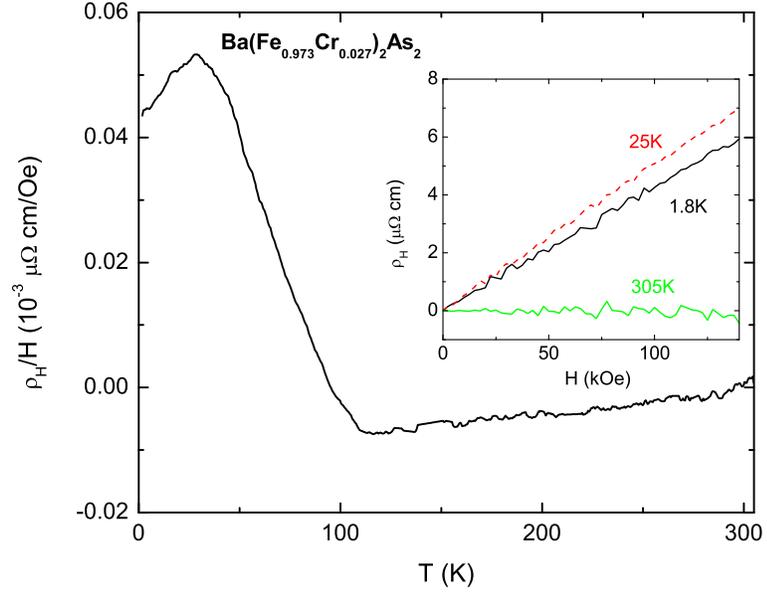}
\end{center}
\caption{(Color online) Temperature dependent  Hall resistivity ($\rho_H/H$) for $H \| c$. Inset shows field-dependent Hall resistivity.}\label{F1c}
\end{figure}

\begin{figure}
\begin{center}
\includegraphics[angle=0,width=120mm]{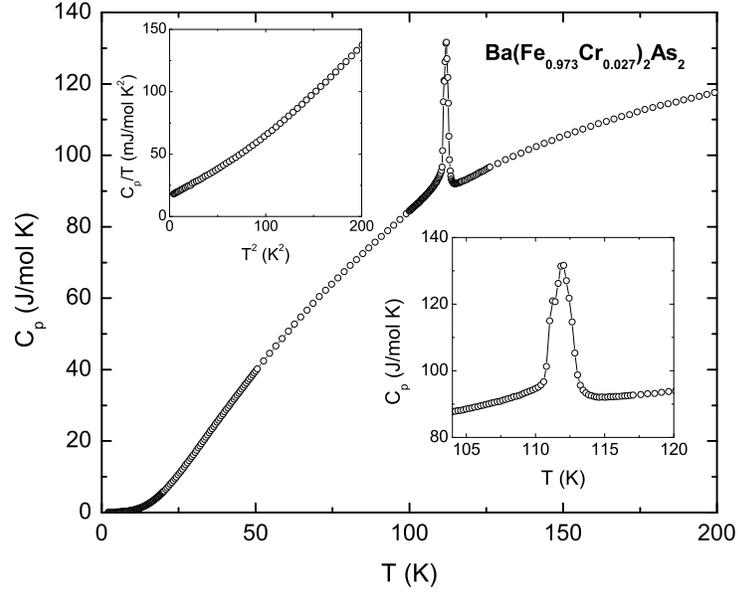}
\end{center}
\caption{Temperature dependent  heat capacity for Ba(Fe$_{0.973}$Cr$_{0.027}$)$_2$As$_2$ single crystals.  Insets show low temperature heat capacity plotted as $C_p/T$ vs. $T^2$ (left) and enlarged transition region (right). }\label{F1d}
\end{figure}

Figs. \ref{F1a}-\ref{F1d} present resistivity, susceptibility, Hall resistivity and heat capacity data for Ba(Fe$_{0.973}$Cr$_{0.027}$)$_2$As$_2$. The structural/magnetic transition temperature for Ba(Fe$_{0.973}$Cr$_{0.027}$)$_2$As$_2$, $T_{sm} \approx 112$ K, is slightly lower than reported \cite{sef09a} for Ba(Fe$_{0.98}$Cr$_{0.02}$)$_2$As$_2$, consistent with slightly higher Cr-doping of the former and is clearly seen in all measurements. The temperature dependent magnetic susceptibility is weakly anisotropic with $\chi_{ab}/\chi_c \approx 1.2$ at 300 K and smaller below $T_{sm}$. This change is primarily due to the fact that the step-like feature at $T_{sm}$ is $\sim 4 - 5$ times larger in $\chi_{ab}$ than in $\chi_c$ (Fig. \ref{F1a}). The slight upturn of the susceptibility at low temperatures for both directions of the applied field might be caused by small amounts of paramagnetic impurities. The temperature dependent electrical resistivity (Fig. \ref{F1b}) manifests a sharp increase upon cooling through $T_{sm}$ and the hysteresis at $T_{sm}$ is at the edge of our resolution $\sim 0.1$ K.  The magnetoresistance (inset) is very small at all measured temperatures. The temperature-dependent Hall resistivity, $\rho_H/H$, (Fig. \ref{F1c}) is small and negative above $T_{sm}$, and then starts to increase rapidly below $T_{sm}$. The field dependence of $\rho_H$ is close to linear over the whole measured temperature range (see inset for representative temperatures). This evolution of the Hall resistivity with temperature is different from that reported for Ba(Fe$_{0.98}$Cr$_{0.02}$)$_2$As$_2$ in Ref. \onlinecite{sef09a}, but is similar to the temperature dependence of the next higher Cr-concentration, Ba(Fe$_{0.96}$Cr$_{0.04}$)$_2$As$_2$,  as well as other hole-doped $AE$Fe$_2$As$_2$ like (Ba$_{0.96}$K$_{0.04}$)Fe$_2$As$_2$. \cite{fan09a} Temperature-dependent specific heat data (Fig. \ref{F1d}) show a single, sharp magnetic/structural transition without a high-temperature knee and the electronic specific heat coefficient (upper inset) is $\gamma \approx 18$ mJ/mol K$^2$. Generally speaking, in many aspects the above data are similar to those reported in Ref. \onlinecite{sef09a}.

\begin{figure}
\begin{center}
\includegraphics[angle=0,width=120mm]{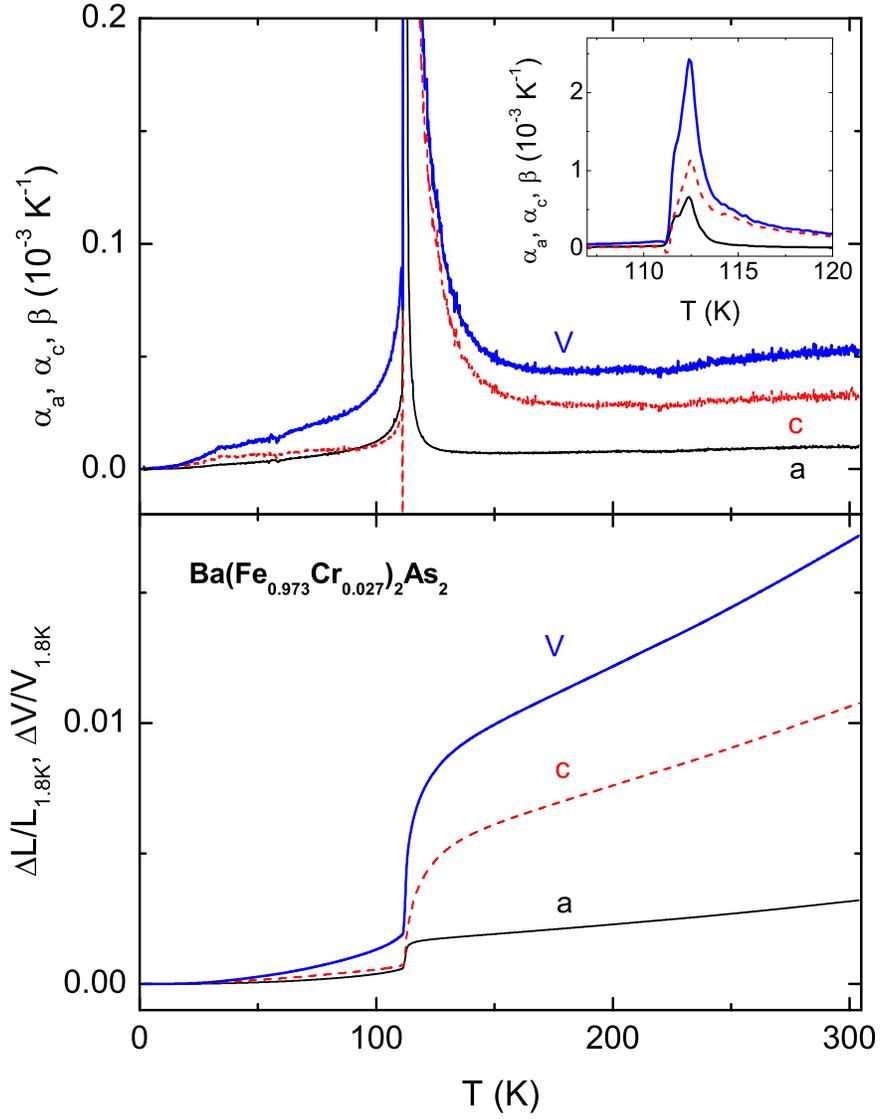}
\end{center}
\caption{(Color online) Anisotropic, temperature dependent thermal expansivity (lower panel) and thermal expansion coefficient (upper panel) of Ba(Fe$_{0.973}$Cr$_{0.027}$)$_2$As$_2$. Inset to the upper panel shows the thermal expansion coefficient near $T_{sm}$.}\label{F2}
\end{figure}

The temperature-dependent, anisotropic, thermal expansivity and thermal expansion coefficients are shown in Fig. \ref{F2}. The structural/magnetic phase transition is sharp. The thermal expansion coefficients above the transition are positive and similar to those measured for pure BaFe$_2$As$_2$. \cite{bud09a} The step-like feature at the transition is larger in the $c$-axis thermal expansivity than in the $a$-axis one, whereas the relative changes in the $a$- and $c$- axes between 119 K and 100 K in Ref. \onlinecite{sef09a} appear to be similar, and the average high temperature $a$-axis thermal expansion in the above work also appears to be negative. We note, however, that the "bulk" thermal expansion measurements yield an average thermal expansion and are not sensitive to possible change in structural symmetry in different phases.

Two, more subtle, observations can be made by examining aforementioned data. Firstly, in heat capacity and thermal expansion (see insets to Figs. \ref{F1d} and \ref{F2}) as well as in the deriative of the temperature dependent resistivity, $d\rho/dT$ (not shown here), it appears that the transition is split in two, spaced by $\sim 1$ K, similarly to the split structural and magnetic transitions in Ba(Fe$_{1-x}$$TM$$_{x}$)$_2$As$_2$ ($TM$ - transition metal). \cite{nin08c,chu09a,pra09a,les09a,bud09a,can09a,nin09a} Secondly, a rather broad anomaly / crossover can be seen in magnetic susceptibility, resistivity, Hall resistivity and thermal expansion (Figs. \ref{F1a}-\ref{F1c}, \ref{F2}) at approximately 30 - 35 K. The origin of this feature is not clear at this point and may warrant further studies.

\begin{figure}
\begin{center}
\includegraphics[angle=0,width=80mm]{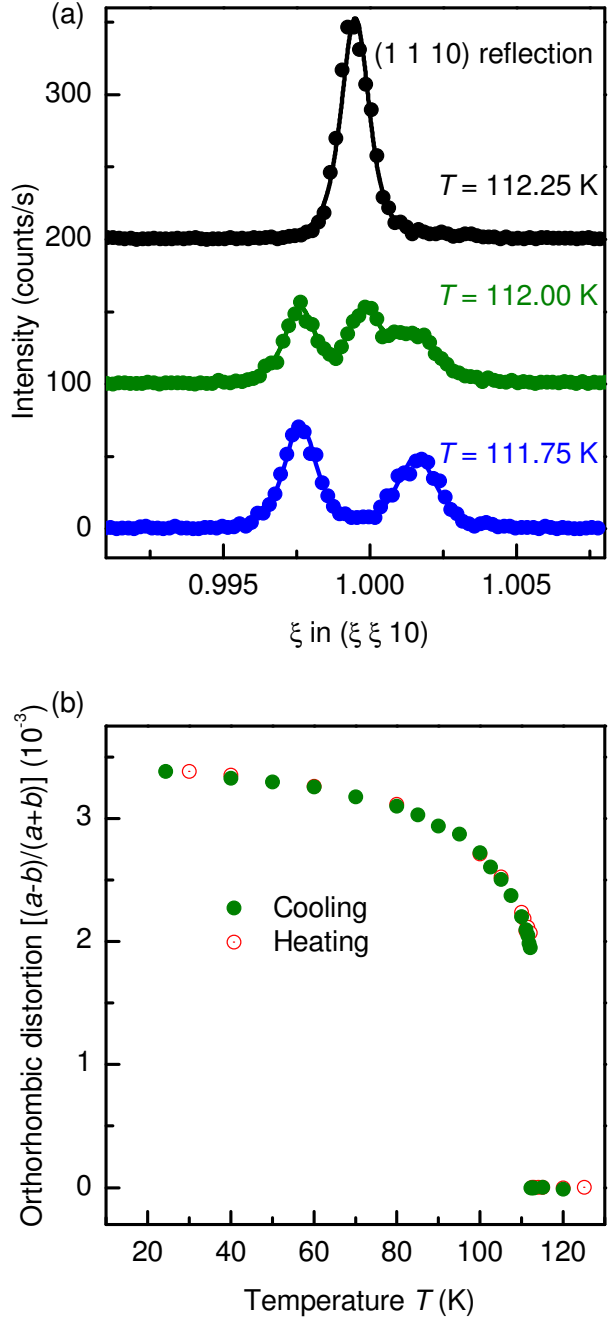}
\end{center}
\caption{(Color online) (a) $(\xi~\xi~0)$ scans through the position of the tetragonal $(1~1~10)$ reflection for temperatures close to the tetragonal-to-orthorhombic transition and for decreasing temperatures.  The offset between every data set is 100 counts/s. The lines represent fit to the data to obtain the reflection positions and corresponding orthorhombic splitting, $(a-b)/(a+b)$, shown in (b). In (b), close (green) and open (red) circles represent orthorhombic splitting during decreasing and increasing temperature scans, respectively.  The error bar for the orthorhombic splitting is less than the symbol size and not shown.}\label{F3}
\end{figure}

Figure \ref{F3} summarizes the temperature dependent, single crystal x-ray diffraction data collected on Ba(Fe$_{0.973}$Cr$_{0.027}$)$_2$As$_2$. Figure \ref{F3}(a) shows the evolution of the $(1~1~10)$ reflection as the sample is cooled through $T_{sm} \approx 112$ K. Whereas there is a clear splitting
in the $(1~1~10)$ reflection in $(\xi~\xi~0)$ scans below 112 K, no change in the shape of the $(0~0~10)$ reflection between 25 K and 125 K was observed. This is consistent with a tetragonal-to-orthorhombic phase transition, from space group $I4/mmm$ to $Fmmm$, with
a distortion along the $(1~1~0)$ direction, as observed in the parent BaFe$_2$As$_2$ compound as well as for other $AE$Fe$_2$As$_2$ compounds. \cite{rot08a,yan08a,nin08a}  Figure \ref{F3}(a) also shows that there is a narrow temperature range ($\leq 0.5$ K) where coexistence between the higher temperature tetragonal phase and the lower temperature orthorhombic structure was observed.  Figure \ref{F3}(b) plots the temperature dependence of the orthorhombic distortion. Below $T_{sm} \approx 112$ K there is an abrupt jump in the orthorhombicity (also evident in Fig. \ref{F3}(a)) which then continues to evolve as the temperature is lowered further.  The abrupt nature of the transition at $T_{sm}$ together with the finite range of coexistence between the high and low temperature structures argues strongly for a first order structural transition.
\\

The splitting we observe at 100 K (the lowest temperature shown in Fig. 1(b) of Ref. \onlinecite{sef09a}) is approximately 0.030 \AA.  This is consistent with the general trend of reducing the orthorhombic splitting at $T_{sm}$ when it is suppressed by doping \cite{nin08b,suy09a} (Rotter et al. observed a 0.038 \AA~splitting in pure BaFe$_2$As$_2$ at 100 K. \cite{rot08b})  It should be noted that in Ref. \onlinecite{sef09a} the splitting reported for pure BaFe$_2$As$_2$ is a significantly smaller, $\sim 0.015$ \AA.  Given that (i) our Cr doping level is slightly higher than the $0.02 \pm 0.01$ reported in Ref. \onlinecite{sef09a} and (ii) there is a clear tetragonal to orthorhombic, structural phase transition seen in pure BaFe$_2$As$_2$ and Ba(Fe$_{0.973}$Cr$_{0.027}$)$_2$As$_2$, it is unlikely that there is a tetragonal to tetragonal phase transition in Ba(Fe$_{0.98}$Cr$_{0.02}$)$_2$As$_2$.

In summary, thermodynamic, structural, and transport measurements on Ba(Fe$_{0.973}$Cr$_{0.027}$)$_2$As$_2$ single crystals show sharp anomalies at $T_{sm} \approx 112$ K associated with a structural/magnetic phase transition. Single crystal x-ray diffraction measurements unambiguously identified the structural transition from the high-temperature tetragonal
$I4/mmm$ to the low-temperature orthorhombic $Fmmm$ structure as being first order. So, in contrast to the earlier report \cite{sef09a} the nature of the structural transition appears to be robust to small doping levels for different types of doping.

\begin{acknowledgments}

Work at the Ames Laboratory was supported by the US Department of Energy - Basic Energy Sciences under Contract No. DE-AC02-07CH11358. We thank R. J. McQueeney for useful comments. SLB and PCC both acknowledge M. T. C. Apoo for providing important insight into this problem.

\end{acknowledgments}

\clearpage

\clearpage

\clearpage

\clearpage

\clearpage

\clearpage

\end{document}